\documentclass[aps,pre,preprint,superscriptaddress]{revtex4-1}

\usepackage{graphicx}
\usepackage{dcolumn}
\usepackage{bm}

\begin{document}


\title{Insight of breaking of powerful axisymmetrically-polarized laser pulses in under-dense plasma}


\author{Nobuhiko Nakanii}
\email[]{nobu.nakanii@ppc.osaka-u.ac.jp}
\affiliation{Photon Pioneers Center, Osaka University}
\affiliation{CREST, Japan Science and Technology Agency}
\author{Tomonao Hosokai}
\affiliation{Graduate School of Engineering, Osaka University}
\affiliation{Photon Pioneers Center, Osaka University}
\affiliation{CREST, Japan Science and Technology Agency}
\author{Naveen C. Pathak}
\affiliation{Photon Pioneers Center, Osaka University}
\author{Shinichi Masuda}
\affiliation{Photon Pioneers Center, Osaka University}
\affiliation{CREST, Japan Science and Technology Agency}
\author{Alexei G. Zhidkov}
\affiliation{Photon Pioneers Center, Osaka University}
\affiliation{CREST, Japan Science and Technology Agency}
\author{Hiroki Nakahara}
\affiliation{Graduate School of Engineering, Osaka University}
\author{Kenta Iwasa}
\affiliation{Graduate School of Engineering, Osaka University}
\author{Yoshio Mizuta}
\affiliation{Graduate School of Engineering, Osaka University}
\author{Naoki Takeguchi}
\affiliation{Graduate School of Engineering, Osaka University}
\author{Takamitsu P. Otuka}
\affiliation{Photon Pioneers Center, Osaka University}
\author{Keiichi Sueda}
\affiliation{Photon Pioneers Center, Osaka University}
\author{Hirotaka Nakamura}
\affiliation{Graduate School of Engineering, Osaka University}
\author{Michiaki Mori}
\affiliation{Kansai Photon Science Institute, Japan Atomic Energy Agency}
\author{Masaki Kando}
\affiliation{Kansai Photon Science Institute, Japan Atomic Energy Agency}
\author{Ryosuke Kodama}
\affiliation{Graduate School of Engineering, Osaka University}
\affiliation{Photon Pioneers Center, Osaka University}
\affiliation{Institute of Laser Engineering, Osaka University}


\date{\today}

\begin{abstract}
Interaction of axisymmetrically-polarized (radially or azimuthally-polarized), relativistically intense laser pulses (ALP) with under-dense plasma is shown experimentally to be different from the interaction of conventional Gaussian pulses. The difference is clearly observed in distinct spectra of scattered laser light as well as in appearance of a strong side emission of second harmonic in the vicinity of focus spot. According 3D particle-in-cell simulations, this is a result of instability in the propagation of ALP in under-dense plasma. Laser wakefield acceleration of electrons by ALP, therefore, is less efficient than that by Gaussian laser pulses but ALP may be interesting for efficient electron self-injection.
\end{abstract}


\maketitle

Even though laser-driven plasma acceleration of electrons gradually evolves from a theory to a technology ready for applications in past a few decades [1-9] many scientific questions have yet to be solved. Among them are solutions for stability and for control of beam parameters such as beam energy and duration, energy spread and total charge, jitter and emittance. A solution should include two main stages of laser wakefield acceleration: electron injection in the acceleration or deceleration part of laser wakefield and acceleration itself [10-12]. 
  
Quality of laser filed distribution in the focus spot is a key part of any laser technique. Usually, for laser wakefield acceleration (LWFA) Gaussian-shape laser pulses are used. Propagation of Gaussian laser pulses in under-dense plasma was a matter of study for decades. Self-focusing, filamentation, hosing, instabilities, breaking of pulse wake are well investigated for high power short laser pulses with the Gaussian profile [13-15]. However, some laser modes may be useful for controllable electron self-injection, further electron acceleration and/or probing of quality of plasma optics. Here, axisymmetrically-polarized, radially ($E_r$, $B_\phi$) or azimuthally-polarized ($E_\phi$, $B_r$), laser pulses are of particular interest being two TEM01 or TEM10 modes, the nearest to Gaussian (TEM00) mode. Such modes tightly focused have been proposed theoretically for the electron vacuum acceleration [16-18] and for efficient electron/positron wakefield acceleration [19]. To our knowledge, effects of propagation such pulses in under-dense plasma have yet to be examined both theoretically and experimentally. 

Intensity of axisymmetrically-polarized laser pulses (ALP) has donut-shape in a pulse focal spot. The transverse ponderomotive force of the ALP, acting towards the laser axis, can collect large number of electrons into the central limited area of the spot [19, 20]. The increase of charge density of electron beam can be expected due to injecting these massive electrons to wake-field excited by another laser pulse in staging acceleration, on the other hand, the electron compression at the laser axis can cause pulse instability resulting in pulse filamentation [20]. In this letter we report on experiments of laser wake-field acceleration (LWFA) with using the ALP along with their support by three dimensional particle-in-cell simulation [20].

Experimental setup is shown in Fig. 1(a). This experiment has been performed with a 40-TW Ti:Sapphire laser system (Amplitude Technologies) at Photon Pioneers Center, Osaka University based on a chirped pulse amplification (CPA) technique. The pulse energy on target was varied up to 660 mJ and the pulse duration was 50 fs. The central wavelength of the laser pulse is 800 nm. The contrast ratio between the main pulse and the nanosecond pre-pulse caused by the amplified spontaneous emission (ASE) is $\sim$ 10$^{10}$, which is measured by third-order cross correlator (SEQOIA, Amplitude Technology). The maximum laser intensity of the ALP on the target is estimated to be 3 $\times$ 10$^{19}$ W/cm$^2$. A linearly-polarized(horizontally-polarized) laser pulse (LLP) with diameter of $\sim$ 50 mm is converted to ALP by passing through an 8-divided wave-plate [21, 22]. As shown in Fig.1 (b), the 8-divided wave-plate is consisted from 8 pieces of 1/2 $\lambda$ wave-plate with different optical axis.　Rotating the wave-plate enables to switch radial, azimuth and the spiral polarizations between these polarizations. The ALP is focused on 100 $\mu$m from the front edge of the slit nozzle by a gold-coated off-axis parabolic mirror (OAP) with f/3.6 (f = 177.8 mm). The focal spot patterns in vacuum with/without using the 8-divided waveplate on target measured by focal spot monitor are shown in Fig.1(c) and (d), respectively. The focal spot of the ALP has clearly donut shape. 

\begin{figure}
\includegraphics[width=12cm]{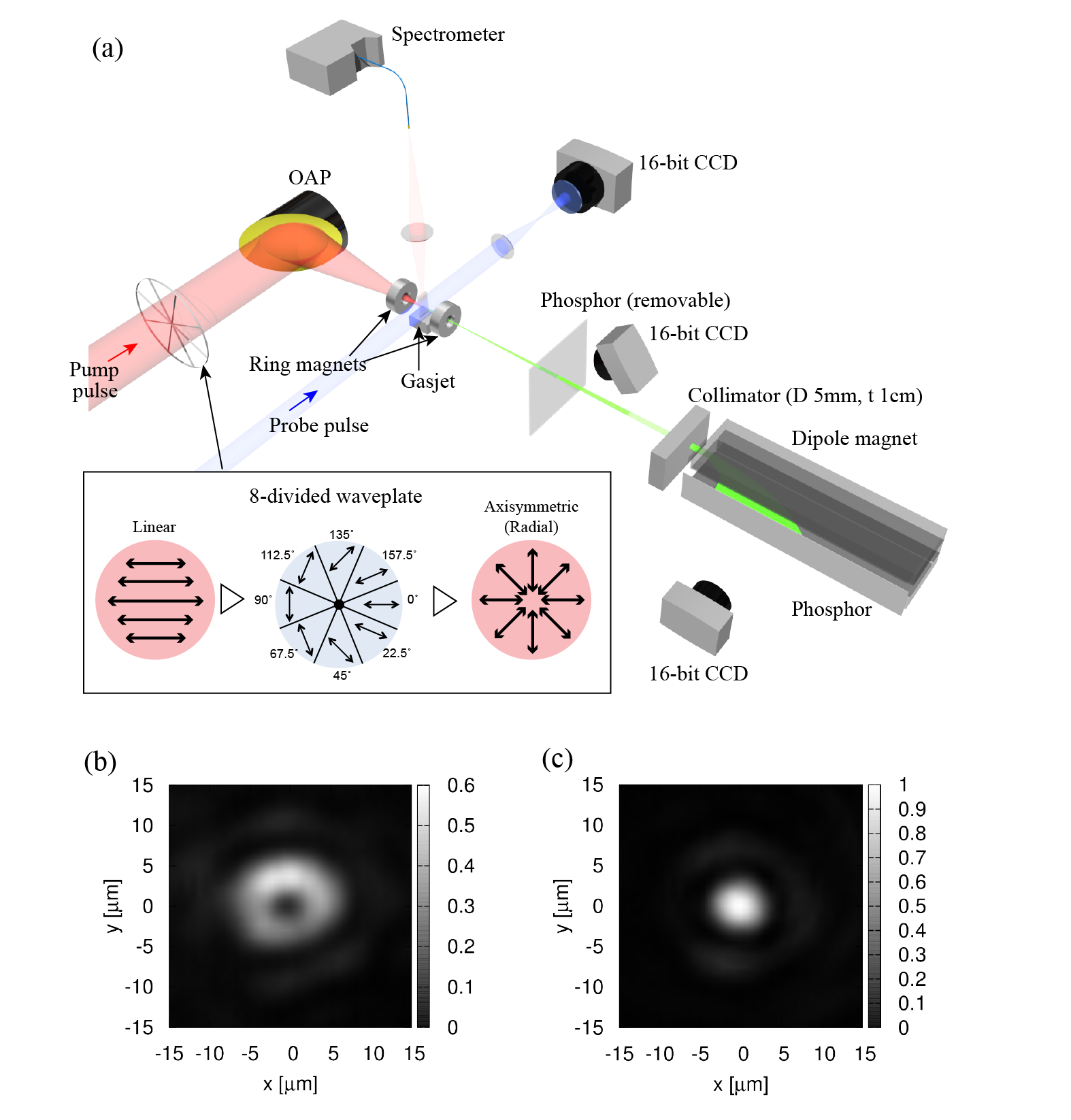}
\caption{\label{fig1} (a) Experimental schematic (b) Focus spot profile of ALP with the 8-divided waveplate and (c) LLP without the waveplate.}
\end{figure}

The length of gas jet was 1.2 mm length. The gas density is estimated to be about 3.4 $\times$ 10$^{19}$ cm$^{-3}$. To stabilized the pointing and the profile of the beam by plasma micro optics, a magnetic device which consists of two ring-shaped neodymium magnets applies external magnetic field (B $\sim$ 0.25 T) in laser propagation direction [23-25]. The spatial distribution of ejected electron beams was measured by a phosphor screen (Mitsubishi Chemical Co. LTD, DRZ-High) with diameter of 9 cm. The screen is located at 43 cm away from the gas jet target. The DRZ-High screen is sensitive to high-energy particles and radiations, so that the front side of the screen is laminated with a 12-$\mu$m-thick aluminum foil to avoid exposure to the laser pulses, scattering lights and low-energy electrons. An electron spectrometer using a dipole magnet measures the energy spectrum of the beams. The energy-resolved electrons turned by the dipole magnet (B $\sim$ 0.25 T) irradiates to another phosphor screen. A 1-cm-thick aluminum collimator with 5-mm-diameter hole is placed in front of the electron spectrometer. The scintillating images on these phosphor screens made by the deposited electrons are recorded by charged coupled device (CCD) cameras (Bitran Co., BU-51LN) with a commercial photographic lens from the backside of the screen. To reduce the background noise due to scattering light of laser pulses, band-pass filters (BG-39, $\lambda$ = 360 - 580 nm) are set at the front of the cameras.

The pulse with a few percent of the laser energy was split by a pellicle and is converted to second harmonic ($\lambda$ $\sim$ 400 nm, $\tau$ $\sim$ 50 fs) by beta-barium borate ($\beta$-BaB$_2$O$_4$, BBO) crystal. This second harmonic is used as probe pulse to observe the condition of plasma by shadowgraph imaging from the side. At the upper side of the gas jet, an optical spectrometer is placed to observe spectrum of the side scattering from the laser plasma interactions. The spectral range of the optical spectrometer is from 520 to 1180 nm. 

Fig. 2 shows spatially and temporally integrated spectra of side scattering to the normal direction of the laser axis caused by interaction of ALP or LLP with gas jets. One can see significant difference between spectra from ALP and LLP, which is a result of inequality in ALP and LLP propagation in under-dense plasma. While the spectrum generated by an LLP is a typical for LWFA reflecting the strong red shift owing to formation of plasma wave in the laser wake [26], the spectrum of the ALP is slightly broadened around the fundamental wavelength. No signature of wake formation is seen for ALP. However it may be effect of exposition: if an ALP decomposes during its propagation the scattering light with the red shift may be too weak for detection. 

\begin{figure}
\includegraphics[width=12cm]{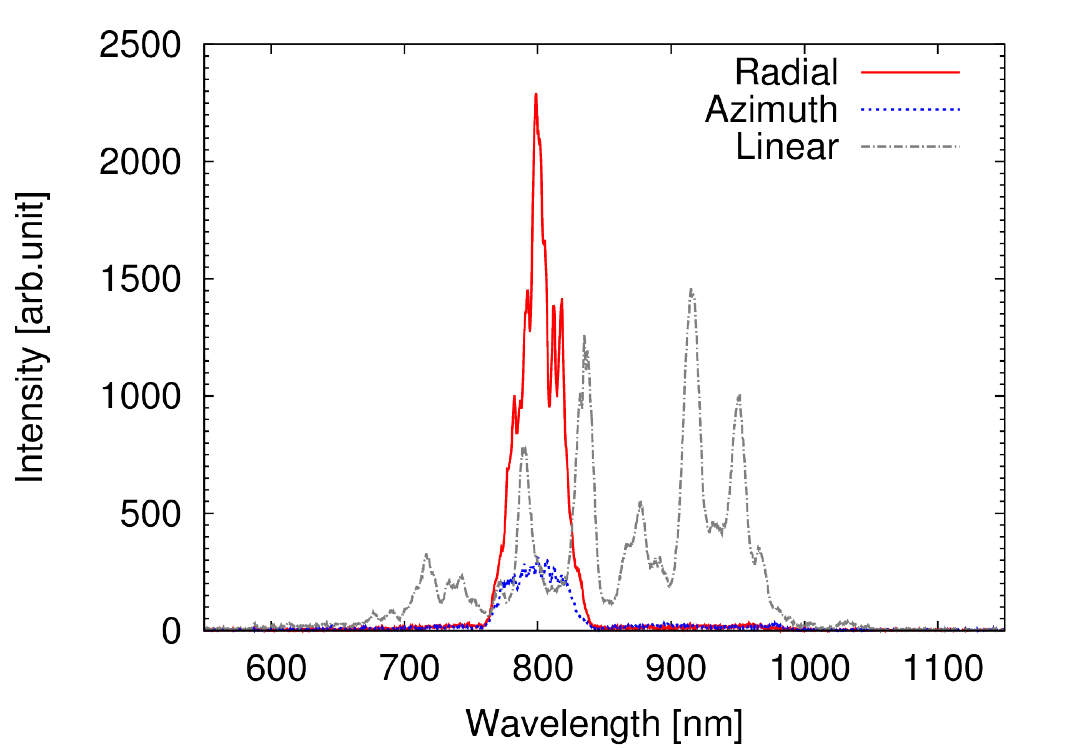}
\caption{\label{fig2} Spectra of side scattering by using the radially, azimuthally and linearly-polarized pulses with pulse energy of 660 mJ.}
\end{figure}

Another sign of ALP decomposition during its propagation is exhibited by Fig.3 where (a) shadowgraph images of the plasmas created by ALP and LLP with a probe pulse ($\sim$ 400 nm) and (b) their profiles after transmission through horizontal polarizer are presented. In the shadowgraph images, one can see strong second harmonic generation (SHG) localized near the focus point of ALP. In contrast to SHG emission reported in Ref. [27], the SHG length in our experiments is quite short about 100 $\mu$m. In the case of LLP we did not observe such SHG emission in all experiments. Since the probe pulse has the vertical linear polarization it cannot be transmitted through horizontal linear polarizer as in Fig. 3(b). Therefore, one can see the strong SHG with different polarization with probe pulse in the cases of the ALP.

\begin{figure}
\includegraphics[width=12cm]{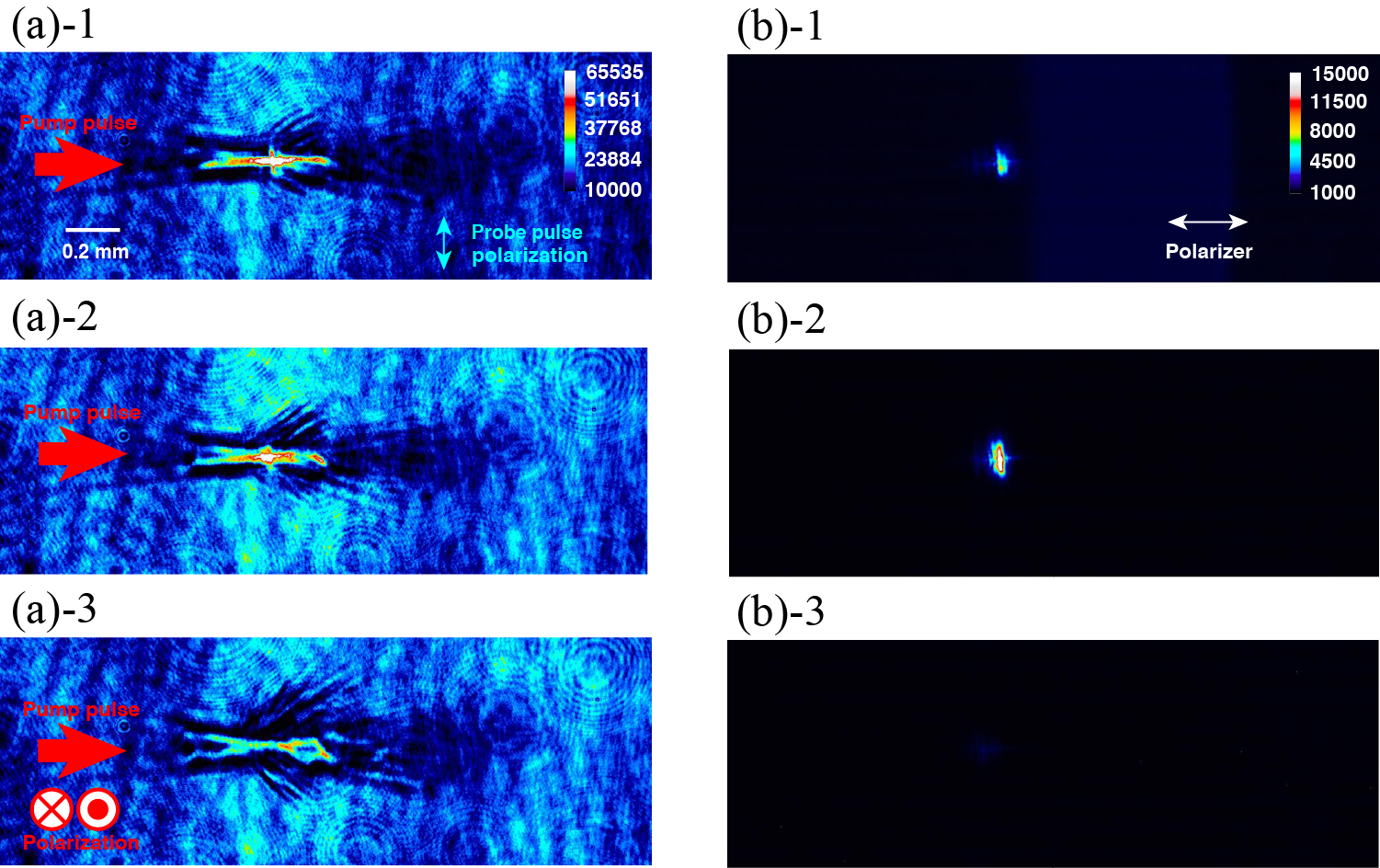}
\caption{\label{fig3} (a) Shadowgraph images of the plasmas created by the ALP and the LLP and (b) their transmission profiles through horizontal polarizer. The numbers 1, 2 and 3 mean radial, azimuth and linear polarizations, respectively. }
\end{figure}

To understand source of strong SHG emission we performed 3D particle-in-cell simulation with parameters close to experimental [details of calculation can be found in Ref [20]]. In Fig. 4(a),(b),(c) results of calculations are given for the intensity of ALP. One can see a clear decomposition (filamentation) of ALP after propagation quite a short distance $\sim$ 300 $\mu$m from the focus spot.  The nature of the instability is discussed in Ref. [20]. To clarify effect of the decomposition on the spectra of side scattered light we perform the spatial analysis for k-vectors for y-component of laser field before (a), during (b), and after (c) decomposition. One can see in Fig. 4(d) that SHG appears in the time of pulse decomposition and then vanish. We attribute this process to Raman scattering of laser light on electrons compressed near the laser axis by the ponderomotive force of ALP. According to our understanding the SHG occurs after the instability develops.

\begin{figure}
\includegraphics[width=12cm]{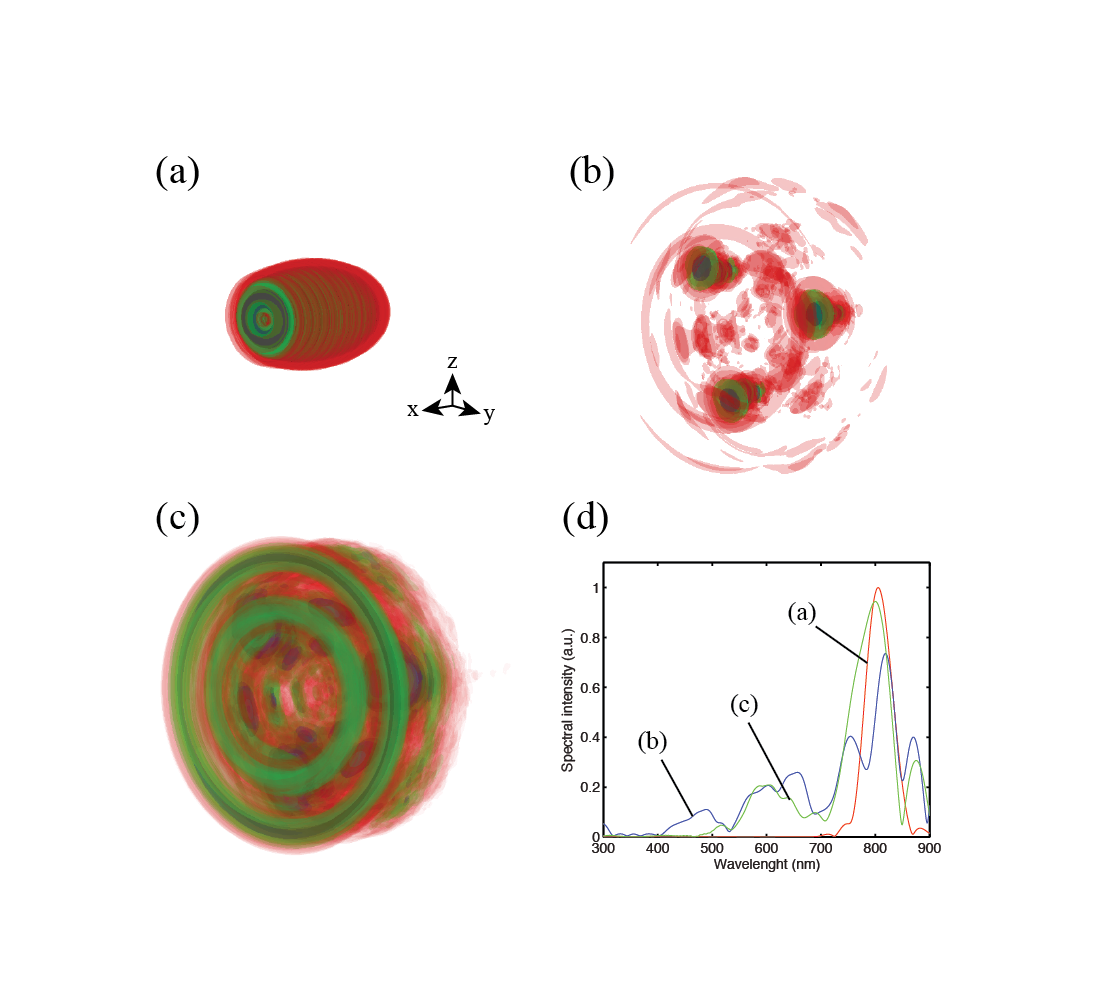}
\caption{\label{fig4} 3D contour plots of intensity evolution of a radially-polarized pulse propagating in uniform underdense plasma[20]. The snap shots show intensity distribution at (a)264 fs, (b) 2.2 ps, and (c) 2.5 ps. (d) The spectrum of the radially-polarized pulse (E$_y$ component) at each timing.}
\end{figure}

Fig. 5 shows spatial profiles of electron beams generated by the ALP (radial and azimuth polarizations) and the LLP with pulse energies 660 mJ on target. As shown in Fig. 5 (a) and (b), the divergence of the electron beam in the case of the ALPs is smaller than the LLP in Fig. 5(c). Energy spectra of the electron beams with each polarization are also shown in Fig. 5(d). The temperature of the electron beam produced by the ALP is lower than the LLP. However, it can be caused by slightly lower intensity of the ALP than the LLP. In the acceleration by ALP, pre-pulse effect may drastically change the acceleration process: according to Ref.[20] there is no ALP decomposition in plasma channels. In pre-plasma formed by an ALP pre-pulse the acceleration may not much different from that by LLP. In the present experiments we worked with a short-focus laser pulse and a short gas jet which length is $\sim$ 1 mm. Therefore in electron spectra we see distributions of self-injected electrons without further acceleration in the case of ALP and not essential acceleration in the case of LLP. Since the electron self-injection occurs mostly in the vicinity of focus spot we observe no significant difference in LWFA with LLP and ALP. An increase of gas jet length may drastically change this result.

\begin{figure}
\includegraphics[width=12cm]{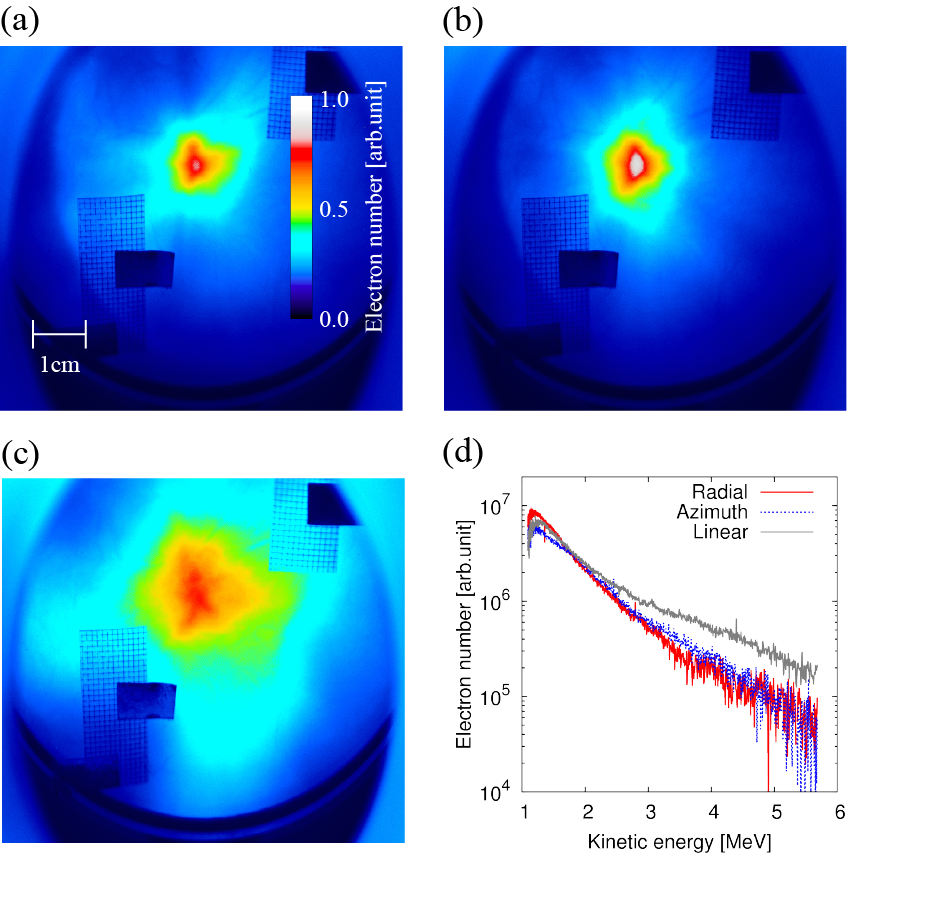}
\caption{\label{fig5} Spatial profiles of electron beams in the case of (a) radial, (b) azimuthal, and (c) linear polarization with pulse energy of 660mJ, and (d) energy spectra of electron beams for each polarization.}
\end{figure}

In conclusion, we have demonstrated experimentally the strong decomposition of axisymmetrically-polarized laser pulses in under-dense plasma. During the decomposition the strong SHG is emitted in quite short distance. In the case of linearly-polarized laser pulses SHG emission does not occur. The SHG confirms the electron compression near the laser axis that result in the pulse instability. Due to pulse decomposition no significant red shift is observed in time- and space-integrated spectra of scattering light. Nevertheless laser wakefield acceleration of plasma electrons up to few MeV energies has been observed. It is less efficient than the electron acceleration by a linearly-polarized laser pulse. According to Ref.[20] ALP can propagate in deep plasma channels without the decomposition. This means that the ALP can be used for probing the plasma optical elements[28]. For example exit of ALP from a plasma channel will be accompanied by a strong emission of SHG which is easy to detect.

\begin{acknowledgments}
This work was supported by Core Research of Evolutional Science and Technology (CREST) of Japan Science and Technology Agency (JST). It was also supported by Genesis Research Institute, Inc. This work was also funded by ImPACT Program of Council for Science, Technology and Innovation (Cabinet Office, Government of Japan). This work was partially supported by JSPS Core-to-Core Program on International Alliance for Material Science in Extreme States with High Power Laser and XFEL. The computational work was carried out at KEK super-computing center, Tsukuba.
\end{acknowledgments}


\end{document}